\documentstyle[twoside]{article}

\catcode`\@=11
\long\def\@makefntext#1{
\protect\noindent \hbox to 3.2pt {\hskip-.9pt  
$^{{\eightrm\@thefnmark}}$\hfil}#1\hfill}		

\def\@makefnmark{\hbox to 0pt{$^{\@thefnmark}$\hss}}	
	
\def\ps@myheadings{\let\@mkboth\@gobbletwo
\def\@oddhead{\hbox{}
\rightmark\hfil\eightrm\thepage}   
\def\@oddfoot{}\def\@evenhead{\eightrm\thepage\hfil
\leftmark\hbox{}}\def\@evenfoot{}
\def\sectionmark##1{}\def\subsectionmark##1{}}

\oddsidemargin=\evensidemargin
\newcounter{sectionc}\newcounter{subsectionc}\newcounter{subsubsectionc}
\renewcommand{\section}[1] {\vspace{12pt}\addtocounter{sectionc}{1} 
\setcounter{subsectionc}{0}\setcounter{subsubsectionc}{0}\noindent 
	{\tenbf\thesectionc. #1}\par\vspace{5pt}}
\renewcommand{\subsection}[1] {\vspace{12pt}\addtocounter{subsectionc}{1} 
	\setcounter{subsubsectionc}{0}\noindent 
	{\bf\thesectionc.\thesubsectionc. {\kern1pt \bfit #1}}\par\vspace{5pt}}
\renewcommand{\subsubsection}[1] {\vspace{12pt}\addtocounter{subsubsectionc}{1}
	\noindent{\tenrm\thesectionc.\thesubsectionc.\thesubsubsectionc.
	{\kern1pt \tenit #1}}\par\vspace{5pt}}
\newcommand{\nonumsection}[1] {\vspace{12pt}\noindent{\tenbf #1}
	\par\vspace{5pt}}

\newcounter{appendixc}
\newcounter{subappendixc}[appendixc]
\newcounter{subsubappendixc}[subappendixc]
\renewcommand{\thesubappendixc}{\Alph{appendixc}.\arabic{subappendixc}}
\renewcommand{\thesubsubappendixc}
	{\Alph{appendixc}.\arabic{subappendixc}.\arabic{subsubappendixc}}

\renewcommand{\appendix}[1] {\vspace{12pt}
        \refstepcounter{appendixc}
        \setcounter{figure}{0}
        \setcounter{table}{0}
        \setcounter{lemma}{0}
        \setcounter{theorem}{0}
        \setcounter{corollary}{0}
        \setcounter{definition}{0}
        \setcounter{equation}{0}
        \renewcommand{\thefigure}{\Alph{appendixc}.\arabic{figure}}
        \renewcommand{\thetable}{\Alph{appendixc}.\arabic{table}}
        \renewcommand{\theappendixc}{\Alph{appendixc}}
        \renewcommand{\thelemma}{\Alph{appendixc}.\arabic{lemma}}
        \renewcommand{\thetheorem}{\Alph{appendixc}.\arabic{theorem}}
        \renewcommand{\thedefinition}{\Alph{appendixc}.\arabic{definition}}
        \renewcommand{\thecorollary}{\Alph{appendixc}.\arabic{corollary}}
        \renewcommand{\theequation}{\Alph{appendixc}.\arabic{equation}}
        \noindent{\tenbf Appendix \theappendixc #1}\par\vspace{5pt}}
\newcommand{\subappendix}[1] {\vspace{12pt}
        \refstepcounter{subappendixc}
        \noindent{\bf Appendix \thesubappendixc. {\kern1pt \bfit #1}}
	\par\vspace{5pt}}
\newcommand{\subsubappendix}[1] {\vspace{12pt}
        \refstepcounter{subsubappendixc}
        \noindent{\rm Appendix \thesubsubappendixc. {\kern1pt \tenit #1}}
	\par\vspace{5pt}}

\newcommand{\textlineskip}{\baselineskip=13pt}
\newcommand{\smalllineskip}{\baselineskip=10pt}

\def\eightcirc{
\begin{picture}(0,0)
\put(4.4,1.8){\circle{6.5}}
\end{picture}}
\def\eightcopyright{\eightcirc\kern2.7pt\hbox{\eightrm c}} 

\newcommand{\copyrightheading}[1]
	{\vspace*{-2.5cm}\smalllineskip{\flushleft
	{\footnotesize International Journal of Modern Physics A, #1}\\
	{\footnotesize $\eightcopyright$\, World Scientific Publishing
	 Company}\\
	 }}

\def\abstracts#1#2#3{{
	\centering{\begin{minipage}{4.5in}\baselineskip=10pt\footnotesize
	\parindent=0pt #1\par 
	\parindent=15pt #2\par
	\parindent=15pt #3
	\end{minipage}}\par}} 

\newcommand{\bibit}{\nineit}

\renewenvironment{thebibliography}[1]
	{\frenchspacing
	 \ninerm\baselineskip=11pt
	 \begin{list}{\arabic{enumi}.}
	{\usecounter{enumi}\setlength{\parsep}{0pt}
	 \setlength{\leftmargin 12.7pt}{\rightmargin 0pt} 
	 \setlength{\itemsep}{0pt} \settowidth
	{\labelwidth}{#1.}\sloppy}}{\end{list}}

\newcounter{itemlistc}
\newcounter{romanlistc}
\newcounter{alphlistc}
\newcounter{arabiclistc}

\newcommand{\fcaption}[1]{
        \refstepcounter{figure}
        \setbox\@tempboxa = \hbox{\footnotesize Fig.~\thefigure. #1}
        \ifdim \wd\@tempboxa > 5in
           {\begin{center}
        \parbox{5in}{\footnotesize\smalllineskip Fig.~\thefigure. #1}
            \end{center}}
        \else
             {\begin{center}
             {\footnotesize Fig.~\thefigure. #1}
              \end{center}}
        \fi}

\newcommand{\tcaption}[1]{
        \refstepcounter{table}
        \setbox\@tempboxa = \hbox{\footnotesize Table~\thetable. #1}
        \ifdim \wd\@tempboxa > 5in
           {\begin{center}
        \parbox{5in}{\footnotesize\smalllineskip Table~\thetable. #1}
            \end{center}}
        \else
             {\begin{center}
             {\footnotesize Table~\thetable. #1}
              \end{center}}
        \fi}

\def\@citex[#1]#2{\if@filesw\immediate\write\@auxout
	{\string\citation{#2}}\fi
\def\@citea{}\@cite{\@for\@citeb:=#2\do
	{\@citea\def\@citea{,}\@ifundefined
	{b@\@citeb}{{\bf ?}\@warning
	{Citation `\@citeb' on page \thepage \space undefined}}
	{\csname b@\@citeb\endcsname}}}{#1}}

\newif\if@cghi
\def\cite{\@cghitrue\@ifnextchar [{\@tempswatrue
	\@citex}{\@tempswafalse\@citex[]}}
\def\citelow{\@cghifalse\@ifnextchar [{\@tempswatrue
	\@citex}{\@tempswafalse\@citex[]}}
\def\@cite#1#2{{$\null^{#1}$\if@tempswa\typeout
	{IJCGA warning: optional citation argument 
	ignored: `#2'} \fi}}

\def\pmb#1{\setbox0=\hbox{#1}
	\kern-.025em\copy0\kern-\wd0
	\kern.05em\copy0\kern-\wd0
	\kern-.025em\raise.0433em\box0}

\def\fnt#1#2{\footnotetext{\kern-.3em
	{$^{\mbox{\scriptsize #1}}$}{#2}}}

\def\fpage#1{\begingroup
\voffset=.3in
\thispagestyle{empty}\begin{table}[b]\centerline{\footnotesize #1}
	\end{table}\endgroup}

\def\runninghead#1#2{\pagestyle{myheadings}
\markboth{{\protect\footnotesize\it{\quad #1}}\hfill}
{\hfill{\protect\footnotesize\it{#2\quad}}}}
\font\tenrm=cmr10
\font\tenit=cmti10 
\font\tenbf=cmbx10
\font\bfit=cmbxti10 at 10pt
\font\ninerm=cmr9
\font\nineit=cmti9

\font\eightrm=cmr8

\def\qed{\hbox{${\vcenter{\vbox{			
   \hrule height 0.4pt\hbox{\vrule width 0.4pt height 6pt
   \kern5pt\vrule width 0.4pt}\hrule height 0.4pt}}}$}}

\input epsf.tex
\def\DESepsf(#1 width #2){\epsfxsize=#2 \epsfbox{#1}}
\begin{document}

\runninghead{Neutralino Proton Cross Sections in $\ldots$} {
Neutralino Proton Cross Sections in $\ldots$}

\normalsize\textlineskip
\thispagestyle{empty}
\setcounter{page}{1}

\copyrightheading{}			

\vspace*{0.88truein}

\fpage{1}
\centerline{\bf Neutralino Proton Cross Sections in }
\vspace*{0.035truein}
\centerline{\bf SUGRA and D-BRANE
             Models}
\vspace*{0.37truein}
\centerline{\footnotesize R. Arnowitt, B. Dutta and Y. Santoso}
\vspace*{0.015truein}
\centerline{\footnotesize\it Department of Physics, Texas A\&M University}
\baselineskip=10pt
\centerline{\footnotesize\it College Station, TX 77843-4242,
USA}

\vspace*{0.21truein}
\abstracts{We calculate the spin independent
neutralino-proton cross section for  universal SUGRA, non universal 
SUGRA and D-brane models with
R-parity invariance. The regions of maximum cross section in these models 
has started to be probed by the current detectors. 
The minimum cross section generally is 
$\stackrel{>}{\sim} 1\times 10^{-(9-10)}$pb and hence will be accessible in the 
 future
detectors, barring special regions of parameter space where it can reduce to
$\simeq 10^{-12}$pb. However, the squarks and gluinos will be heavy 
($\stackrel{>}{\sim}$1 TeV) in the latter case, but still accessible at the LHC.}{}{}

\textlineskip\vspace*{12pt}
We consider here the models based on gravity
mediated supergravity (SUGRA), where the LSP is generally the lightest
neutralino ($\tilde\chi^0_1$). Neutralinos can be detected directly in dark matter 
experiments by their elastic scattering with nuclear targets. For heavy nuclear
targets, such
scattering is dominated by the spin independent part
 and it is possible to extract then the spin independent neutralino -proton cross section,
$\sigma_{\tilde\chi^0_1-p}$. Current experiments (DAMA, CDMS, UKDMC) have
sensitivity to $\tilde\chi^0_1$ for 
\begin{equation}
\sigma_{\tilde\chi^0_1-p}\stackrel{>}{\sim} 1\times 10^{-6} \, {\rm pb}
\end{equation} and future detectors (GENIUS, Cryoarray) plan to achieve a
sensitivity of 
$\sigma_{\tilde\chi^0_1-p}\stackrel{>}{\sim} (10^{-9}-10^{-10}) \, {\rm pb}$.

We calculate the cross section in three models  based on
unification of the gauge coupling constants at $M_G\cong 2\times 10^{16}$ GeV:
(1) Minimal super gravity GUT (mSUGRA) with universal soft breaking at $M_G$,
 (2) Nonuniversal soft breaking models, and (3) D-brane models (based
on type IIB orientifolds\cite{munoz}). In these models, we examine the
 part of the parameter space being
probed by current experiments and the smallest value of
$\sigma_{\tilde\chi^0_1-p}$ the models are  predicting. 

The parameter space of these models is restricted by the following 
constraints\cite{aads}. 
(1) The electroweak symmetry is radiatively broken. 
(2) The neutralino relic density is \cite{aads} 
$0.02\leq \Omega_{\tilde\chi^0_1}h^2\leq 0.25.$ 
(3) We impose the recent collider bounds (LEP\cite{falk1} and 
Tevatron\cite{comm1}).
(4) We also impose the CLEO constraints\cite{cleo} on  BR($b\rightarrow
s\gamma$). We follow the analysis of \cite{ellis2} 
to convert $\tilde\chi^0_1$-quark cross section to $\tilde\chi^0_1-p$
scattering. For this, we use $\sigma _{\pi N}=65$ MeV \cite{mgo},
$\sigma_0=30$ MeV\cite{bottino2} and r$=24.4\pm 1.5$\cite{leutwyler}. 

In mSUGRA model, $(\sigma_{\tilde\chi^0_1-p})_{\rm max}$ arises for large tan$\beta$ and
small $m_{1/2}$. In Fig.1
($\sigma_{\tilde\chi^0_1-p}$)$_{\rm max}$ is plotted vs. $m_{\tilde \chi^0_1}$
for tan$\beta$=20, 30, 40 and  50.  Current
detectors are sampling the parameter space for large tan
$\beta$, small $m_{\tilde \chi^0_1}$ (and also 
small $\Omega_{\tilde\chi^0_1}h^2$)
\begin{equation} \tan\beta\stackrel{>}{\sim}25;\,\,
m_{\tilde\chi^0_1}\stackrel{<}{\sim}90\,{\rm
GeV};\,\,\Omega_{\tilde\chi^0_1}h^2\stackrel{<}{\sim}0.1.
\end{equation}We have used $m_0\leq
1$ TeV, $m_{1/2}\leq 600$ GeV (corresponding to $m_{\tilde g}\leq 1.5$ TeV,
$m_{\tilde \chi^0_1}\leq 240$ GeV), $|A_0/m_0|\leq 5$, and 2$\leq tan\beta\leq$
50. 

The minimum cross section occurs at small tan$\beta$. For
$m_{\tilde \chi^0_1}\stackrel{<}{\sim} 150$ GeV ($m_{1/2}\leq 350$) where no
coannihilation occurs, one finds
\begin{equation}
\sigma_{\tilde\chi^0_1-p}\stackrel{>}{\sim} 4\times 10^{-9} {\rm pb};\,
 m_{\tilde \chi^0_1}\stackrel{<}{\sim} 140 {\rm GeV}
\end{equation} which would be accessible to detectors that are currently being
planned (e.g. GENIUS).
For larger $m_{\tilde \chi^0_1}$, i.e. $m_{1/2}\stackrel{>}{\sim} 150$ GeV the
phenomena of coannihilation can occur in the relic density analysis since the
light stau, $\tilde \tau_1$, (and also $\tilde e_R$, $\tilde \mu_R$) can become
degenerate with the $\tilde\chi^0_1$. The relic density constraint can then be
satisfied in narrow corridor of $m_0$ of width 
$\Delta m_0\stackrel{<}{\sim}25$ GeV\cite{falk}.

 Fig.2 shows
$\sigma_{\tilde\chi^0_1-p}$ in the domain of large $A_0$ and for two values of
tan$\beta$ for $\mu>0$ (sign convention of ISAJET)\cite{adsprep}. 
The smaller tan$\beta$ gives  the lower cross
section for smaller $m_{1/2}$.  In this case we have
 $\sigma_{\tilde\chi^0_1-p}\stackrel{>}{\sim} 1\times 10^{-9} {\rm pb},\,{\rm
for}\,
 m_{1/2} \leq 600 {\rm GeV},\,\mu>0$ and $A_0\leq4 m_{1/2}$.
This is still within the sensitivity range of proposed detectors.

A surprize cancellation can occur in part of the
parameter space for negative  $\mu$ in the coannihilation region which can greatly reduce
$\sigma_{\tilde\chi^0_1-p}$ \cite{ellis}. In Fig.3,  
 the cross section becomes a minimum ($\cong 1\times 10^{-12}$)
  at about
tan$\beta$=10, and then increases again for larger tan$\beta$\cite{adsprep}. 
  In this domain,
$\sigma_{\tilde\chi^0_1-p}$ would not be accessible to any of the  currently
planned detectors. However, mSUGRA  predicts that this could happen for $m_{1/2}=600$ GeV
i.e.
only when the gluino and squarks have masses greater than 1 TeV (and only for a
restricted region of tan$\beta$).

In the SUGRA models with 
nonuniversal soft breaking, we assume that the Higgs and third generation
squark and slepton masses become nonuniversal. It is possible to 
 greatly increase $\sigma_{\tilde\chi^0_1-p}$,  
by a factor of 10-100 compared to the universal
case by a suitable choice of nonuniversality.  
The current detectors  can probe part of the
parameter space for tan$\beta$ as low as
$tan\beta\simeq 4$. 
${\sigma_{\tilde\chi^0_1-p}}_{\rm min}$ occurs (as in mSUGRA) at the lowest
tan$\beta$ and at the largest
$m_{1/2}$ i.e. in the coannihilation region. For $\mu<0$, there can again be a
cancellation of matrix elements reducing
the cross section to $10^{-12}$ pb when $m_{1/2}= 600$ GeV, tan$\beta\simeq$10
in a restricted part of the parameter space. 

For the
 D-brane model, the lowest value of tan$\beta$ that can be examined by the 
 current detectors is fifteen. We have ${\sigma_{\tilde\chi^0_1-p}}_{\rm
 min}\stackrel{>}{\sim}10^{-9}$ pb for
$\mu>0$, while a cancellation allows $\sigma_{\tilde\chi^0_1-p}$ to fall to
$10^{-12}$ pb for $\mu<0$ at tan$\beta\simeq 12$\cite{adsprep} (with again a
gluino/squark mass spectrum in the TeV domain).

This work was supported in part by NSF grant no.
PHY-9722090.

\nonumsection{References}

\begin{figure}[htb]
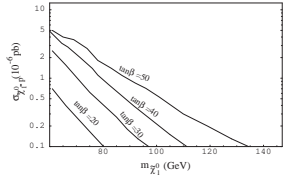

\centerline{ \DESepsf(aads20304050.epsf width 3.70 cm) }
\caption {\label{fig1}  $(\sigma_{\tilde{\chi}_{1}^{0}-p})_{\rm max}$ for mSUGRA
obtained by varying $A_0$ and $m_0$ over the parameter space for  tan${\beta}=20$, 30, 40,
and 50[5]. The relic density constraint has been imposed.}
\end{figure}
\begin{figure}[htb]
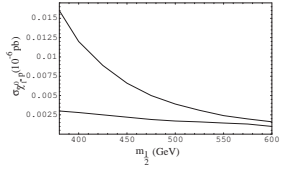

\centerline{ \DESepsf( aadcoan403.epsf width 3.70 cm) }
\caption {\label{fig4} $(\sigma_{\tilde{\chi}_{1}^{0}-p})$ for mSUGRA in the
coannihilation region for   tan${\beta}=40$ (upper curve) and tan${\beta}=3$
(lower curve), 
$A_0=4 m_{1/2}$, $\mu>0$[6].}
\end{figure}
\begin{figure}[htb]
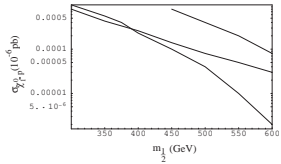

\centerline{ \DESepsf(aadcoan51020.epsf width 3.70 cm) }
\caption {\label{fig5} $(\sigma_{\tilde{\chi}_{1}^{0}-p})$ for mSUGRA and
$\mu<0$ for (from top to bottom on right)   tan${\beta}$=20, 5 and 10. Note that
for tan${\beta}\geq 10$, the curves terminate at the left due to the
$b\rightarrow s\gamma$ constraint[6].}
\end{figure}

\begin{thebibliography}{99}
\bibitem{munoz}L. Ibanez, C. Munoz and S. Rigolin, 
{\bibit Nucl. Phys. B553} (1999) 43; 
M. Brhlik, L. Everett, G. Kane and J. Lykken; {\bibit Phys. Rev. D62} (2000)
 035005.

\bibitem{falk1}J. Ellis, T. Falk, G. Ganis and K.A. Olive,  hep-ph/0004109.

\bibitem{comm1} D0 Collaboration, {\bibit Phys. Rev. Lett. 83} (1999) 
4937.

\bibitem{cleo}M. Alam et al., {\it Phys. Rev. Lett. 74} (1995) 2885.

\bibitem{aads}E. Accomando, R. Arnowitt, B. Dutta and Y. Santoso, 
{\bibit Nucl. Phys. B585} (2000) 124; R. Arnowitt, B. Dutta and Y. Santoso,
 hep-ph/0008336.


\bibitem{ellis2}J. Ellis and  R. Flores,{\bibit  Phys. Lett. B263} (1991) 259; 
{\bibit B300} (1993) 175.

\bibitem{mgo}M. Ollson, hep-ph/0001203; M. Pavan, R. Arndt, I. Stravkovsky,
 and R. Workman, nucl-th/9912034.

\bibitem{bottino2}A. Bottino, F. Donato, N. Fornengo,  and S. Scopel, {\bibit  Astropart.
Phys. 13} (2000) 215.

\bibitem{leutwyler}H. Leutwyler, {\bibit  Phys. Lett. B374} (1996) 163.

\bibitem{falk}J. Ellis, T. Falk, K.A. Olive and  M. Srednicki, 
{\bibit Astropart. Phys. 13} (2000) 181. 
\bibitem{ellis}J. Ellis, A. Ferstl and  K.A. Olive; hep-ph/0001005.
\bibitem{adsprep}R. Arnowitt, B. Dutta, Y. Santoso, in preparation.
\end{thebibliography}
\end{document}